\documentclass[conference]{IEEEtran}
\IEEEoverridecommandlockouts

\usepackage{pgfplots}
\pgfplotsset{width=8cm,compat=1.9}

\usepackage{cite}
\usepackage{amsmath,amssymb,amsfonts}
\usepackage{algorithmic}
\usepackage{graphicx}
\usepackage{textcomp}
\usepackage{xcolor}

\def\BibTeX{{\rm B\kern-.05em{\sc i\kern-.025em b}\kern-.08em
    T\kern-.1667em\lower.7ex\hbox{E}\kern-.125emX}}
\makeatletter
    \newcommand{\linebreakand}{%
      \end{@IEEEauthorhalign}
      \hfill\mbox{}\par
      \mbox{}\hfill\begin{@IEEEauthorhalign}
    }
\makeatother

\def\footnoterule{\relax%
  \kern-5pt
  \hbox to \columnwidth{\hfill\vrule width 0.5\columnwidth height 0.4pt\hfill}
  \kern4.6pt}
\makeatother

\begin{document}

\title{Applying wav2vec2 for Speech Recognition on Bengali Common Voices Dataset\\}

\author{
\IEEEauthorblockN{1\textsuperscript{st} H.A.Z Sameen Shahgir}
\IEEEauthorblockA{\textit{Undergraduate, Computer Science and Engineering} \\
\textit{Bangladesh University of Engineering and Technology}\\
Dhaka, Bangladesh \\
1805053@ugrad.cse.buet.ac.bd}
\and
\IEEEauthorblockN{2\textsuperscript{nd} Khondker Salman Sayeed}
\IEEEauthorblockA{\textit{Undergraduate, Computer Science and Engineering} \\
\textit{Bangladesh University of Engineering and Technology}\\
Dhaka, Bangladesh \\
1805050@ugrad.cse.buet.ac.bd}
\linebreakand
\IEEEauthorblockN{3\textsuperscript{rd} Tanjeem Azwad Zaman}
\IEEEauthorblockA{\textit{Undergraduate, Computer Science and Engineering} \\
\textit{Bangladesh University of Engineering and Technology}\\
Dhaka, Bangladesh \\
1805006@ugrad.cse.buet.ac.bd}

}

\maketitle

\begin{abstract}
Speech is inherently continuous, where discrete words, phonemes and other units are not clearly segmented, and so speech recognition has been an active research problem for decades. In this work we have fine-tuned \textit{wav2vec 2.0} to recognize and transcribe Bengali speech -- training it on the \textit{Bengali Common Voice Speech Dataset}. After training for 71 epochs, on a training set consisting of 36919 mp3 files, we achieved a training loss of 0.3172 and WER of 0.2524 on a validation set of size 7,747.  Using a 5-gram language model, the Levenshtein Distance was 2.6446 on a test set of size 7,747. Then the training set and validation set were combined, shuffled and split into 85-15 ratio. Training for 7 more epochs on this combined dataset yielded an improved Levenshtein Distance of 2.60753 on the test set. Our model was the best performing one on a hidden dataset, achieving a Levenshtein Distance of 6.234\footnote{https://www.kaggle.com/competitions/dlsprint/discussion/349991}, which was 1.1049 units lower than other competing submissions.
\end{abstract}


\section{Introduction}
Speech recognition and transcription is one of the quintessential applications of Machine Learning, with every advancement having wide reaching effects on society as a whole. English speech recognition in particular is at the stage of commercial viability, with products like Alexa, Google Assistant and Siri becoming household names. Unfortunately, the state of the art for Bengali speech recognition is lagging behind, especially considering that it is one of the most widely spoken languages in the world. Several factors such as the lack of commercial incentive, geological clustering of native speakers and a relatively young Bengali IT industry have caused this. 

“LRLs (Low resource Languages) can be understood as less studied, resource scarce, less computerized, less privileged, less commonly taught, or low density, among other denominations.”\cite{LowResourceLangMAC} Bengali can be considered a low-resource language in the sense that, transcribed speech for Bengali (to be used in supervised learning) is very scarce.

Recurrent Neutral Networks (RNN) has long been the go-to solution for sequence to sequence machine translation tasks like Speech Recognition. However, it runs into the problem of vanishing gradients and very high computational cost owing to the lack of parallelizability when fitting long data, such as speech. Each second of audio sampled at 16KHz produces to 16000 elements in its vector representation. Modifications to RNN such as Long short-term memory (LSTM) and Gated Recurrent Unit (GRU) can mitigate, but not quite overcome the limitations of a recurrence based approach.

The Transformer\cite{vaswani2017attention} is a model architecture which eschews recurrence and instead relies entirely on an attention mechanism to draw dependencies between input and output. The Transformer allows for significantly more parallelization and have reached a new state of the art in sequence to sequence translations.

This brings us to Wav2vec2.0, a great match for our requirements of an unsupervised model suited to low resource languages. Wav2vec2.0 is “a framework for self-supervised learning of representations from raw audio data”\cite{wav2vec2OG}  . This model uses a multi-layer convolutional neural network (CNN) to encode speech audio and produce latent speech representations. Spans of these representations are masked and contextualized using a Transformer network. The model then demarcates true latent from distractors through training via contrastive tasks. The model learns discrete speech units as part of this pre-training on unlabeled speech. Afterward, labeled data is used to fine-tune the model with a Connectionist Temporal Classification (CTC) loss. This aids in downstream speech recognition tasks. 

The CNN based speech-to-embedding method applied by the \textit{wav2vec2.0} model is not language dependent, rather it is able to learn the patterns that all spoken languages have in common. As such, pretraining on multiple languages was found to be beneficial even when the target language was not included in the pretraining phase \cite{xu2021simple}. As such, we chose \textit{facebook/wav2vec2-large-xlsr-53}, which was pretrained on about 56 thousand hours of multilingual speech from the MLS, CommonVoice and BABEL datasets \footnote{https://github.com/facebookresearch/fairseq/tree/main/examples/wav2vev}. 

Recently, an addition to the repository of Bengali speech transcription was made by \textit{Bengali Common Voice Speech Dataset}\cite{cvbn}. This dataset allows us to fine-tune Meta's pretrained \textit{wav2vec 2.0} to Bengali to allow better transcription specific to Bengali. In this paper we showcase the training pipeline used to train \textit{wav2vec 2.0} to \textit{Bengali Common Voice Speech Dataset} to achieve a considerable Word Error Rate (WER).

\section{Methodology}
\subsection{Model Selection}
wav2vec2 is currently yields state-of-the-art WER 1.4\% / 2.6\% on the LibriSpeech test/test-other sets\cite{zhang2020pushing}. As such it was a natural choice for Bengali ASR. Two different approaches were considered as the starting point for training, namely a self-supervised pre-trained model (\textit{facebook/wav2vec2-large-xlsr-53})\footnote{https://huggingface.co/facebook/wav2vec2-large-xlsr-53} and an already fine-tuned model (\textit{arijitx/wav2vec2-xls-r-300m-bengali})\footnote{https://huggingface.co/arijitx/wav2vec2-xls-r-300m-bengali} convergent on another similar dataset\cite{kjartansson-etal-sltu2018} (i.e. transfer learning). After preliminary testing, we found that further fine-tuning an already convergent model slightly degrades performance on the target dataset. Our results concur with similar findings where wav2vec2 model was applied to CALLHOME-MA dataset \cite{yi2020applying}.

Therefore, we determined that the self-supervised pretrained model \textit{facebook/wav2vec2-large-xlsr-53} should serve as the basis for further fine-tuning.

\subsection{Preprocessing}
The train split of \textit{Bengali Common Voice Speech Dataset} comprises of 206,951 mp3 files with their corresponding Bengali transcription along with some meta-data such as upvotes, downvotes, gender etc. On quick analysis, the following metrics were found.

\begin{table}[h!]
\centering
\caption{dataset analysis findings}
    \begin{tabular}{|c|c|}
    \hline
       Criterion  & Count  \\ 
       \hline
       $upvotes > downvotes$  & 37405\\
       $upvotes < downvotes$  & 5536\\
       $upvotes = downvotes = 0$ & 161380 \\
       $upvotes > downvotes \text{\ and \ } 1\leq duration \leq 10$ & 36919 \\
       \hline
    \end{tabular}

    \label{table:1}
\end{table}

Since 5536 ($\sim$13 percent) out of 42941 voted data was unreliable, we opted to train using only the subset with  $upvotes > downvotes$.

The mp3 files were then sampled at 16kHz and characters in the label strings were cast to integer hashes using the a vocabulary dictionary forked from the huggingface model  \textit{arijitx/wav2vec2-xls-r-300m-bengali}.

The Transformers\cite{vaswani2017attention} implementation from the Transformers library\footnote{https://huggingface.co/docs/transformers/index} pads each array to match the longest array in the same batch. For faster training, we removed the starting and ending portions of the each sound array if the value at that index was less than some fraction of the maximum value in that array. After testing different values, the cutoff threshold $x<\frac{max(array)}{30}$ was chosen.

Finally, only sound files between 1 and 10 seconds were chosen for training, yielding a final train set of length 36919.

\subsection{Training}

For training we used the pytorch \cite{NEURIPS2019_9015} based implementation of Transformer \cite{vaswani2017attention} model provided and maintained by huggingface.co. We fine-tuned the pretrained \textit{facebook/wav2vec2-large-xlsr-53} which was trained to on unlabelled multilingual speech and intended for further downstream training on labelled data.

First Phase Training was done for a total of 71 epochs, totalling to around 100 hours of training on a single Nvidia K80 GPU. The training run-time was approximately 80 hours for 71 epochs of training. The AdamW Optimizer was used starting with learning rate $5\times10^{-4}$ and weight delay of $2.5\times10^{-6}$. These values were decided on after some preliminary attempts which either failed to converge with larger hyper-parameters or were slow to show appreciable convergence with lower rates.

Around 0.25 WER ($\sim$ Epoch 55), we noticed the training and evaluation metrics plateauing considerably. On 7747 hidden test data, we achieved a Levenshtein Distance of 2.64446. Speech recognition tasks often hinge on exposure to a large vocabulary and since we initially limited ourselves to a subset of the entire dataset, we used a portion of the validation set for further training.

 The main motivation behind two phase training was to increase the model's exposure to new speech and the corresponding text labels. We merged training and validation datasets and applied a 85-15 split. With a weight decay of $2.5\times10^{-9}$ and learning rate $5\times10^{-6}$, we trained for a further 7 epochs, which took a further 9 hours on the Nvidia K80 GPU. The hyper-parameters were significantly lowered to allow prevent the model weights from rapidly deviating.
 
 This produced a slight but noticeable improvement to the Levenshtein Distance, now 2.60753 (down from 2.6446).

\subsection{Post-processing}

Three separate post processing functions were used to generate the final prediction on the hidden test set, namely n-gram Language Model decoding, Unicode normalization and appending a common character to the end of sentences.

Our Language Model of choice was forked from the huggingface model \textit{arijitx/wav2vec2-xls-r-300m-bengali}, generated from IndicCorp language corpus \cite{kakwani2020indicnlpsuite}. The language model improved spelling accuracy and the Levenshtein Distance decreased from 3.30648 to 2.72243 once applied.

Unicode Normalization using \textit{bnunicodenormalizer}\footnote{https://github.com/mnansary/bnUnicodeNormalizer} and appending the "devanagari danda" [Unicode\#2404] to output sentences also resulted in slight improvements.

Since most punctuation was removed during preprocessing, the model outputs didn't contain them. A suitable language model might could possibly reinsert punctuation but we did not explore this possibility in this iteration.

\section{Results}
After a total of about 77 epochs of training, in two phases - our model settled on a Levenshtein Mean Distance Score of 2.60753 on the test set. The training error and evaluation word error metric gradually decreased through out the training process. After training phase 1, the model had a training loss of 0.3172, and an evaluation WER of 0.2524 on the validation set. 

The final model scored a Levenshtein Mean Distance of 6.234 on a new hidden dataset.\footnote{https://www.kaggle.com/competitions/dlsprint/discussion/349991}

\begin{figure}[h!]
    \centering
    \begin{tikzpicture} 
        \begin{axis}[
            xlabel={Epoch},
            ylabel={Training Loss},
            xmin=0, xmax=80,
            ymin=0, ymax=1,
            xtick={10, 20, 30, 40, 50, 60, 70, 80},
            ytick={0, 0.1, 0.2, 0.3, 0.4, 0.5, 0.6, 0.7, 0.8, 0.9, 1.0},
            ymajorgrids=true,
            grid style=dashed,
        ]
        
        \addplot[color=black, smooth, thick] table[x index=0, y index=1, col sep=comma, skip first n=5] {training_data.csv};
        \end{axis}
    \end{tikzpicture}
    \caption{Training Loss vs Epoch} \label{fig:0}
\end{figure}
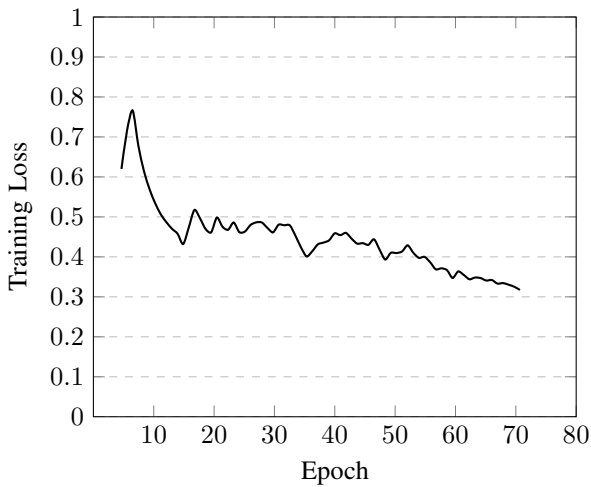

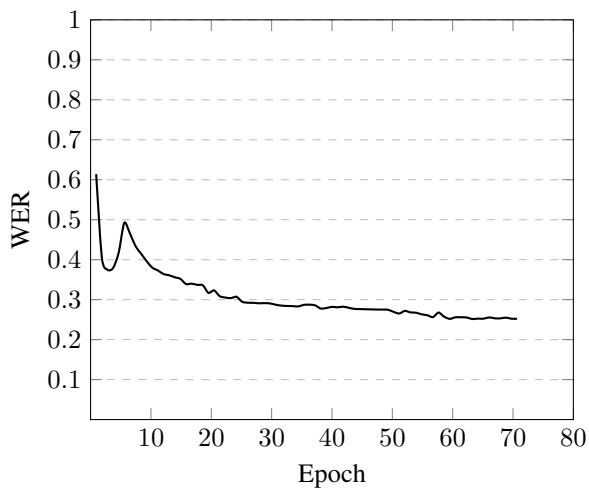
\begin{figure}[h!]
    \centering
    \begin{tikzpicture} \label{plt:1}
        \begin{axis}[
            xlabel={Epoch},
            ylabel={WER},
            xmin=0, xmax=80,
            ymin=0, ymax=1,
            xtick={10, 20, 30, 40, 50, 60, 70, 80},
            ytick={0.1, 0.2, 0.3, 0.4, 0.5, 0.6, 0.7, 0.8, 0.9, 1.0},
            ymajorgrids=true,
            grid style=dashed,
        ]
        
        \addplot[color=black, smooth, thick] table[x=epoch, y=wer, col sep=comma] {training_data.csv};
        \end{axis}
    \end{tikzpicture}
    \caption{WER vs Epoch} \label{fig:1}
\end{figure}

From figure \ref{fig:1} we can see that WER has plateaued at around 60 epochs. Whereas training loss, as in figure \ref{fig:0} showed steady descent.

\begin{table}[h!]
    \centering
       \caption{Training Phase 2 Metrics}
    \begin{tabular}{|c|c|c|}
    \hline
    epoch & Training Loss & Eval.WER\\
    \hline
    1.26   & 0.2552	&	0.1497\\
    2.51   & 0.2482	&	0.1500\\
    3.77   & 0.2497	&	0.1498\\
    5.02   & 0.2474	&	0.1499\\
    6.28   & 0.2493	&	0.1499\\
    \hline
    \end{tabular}
 
    \label{table:2}
\end{table}

\section{Comparison with other works}
This work was the champion on the kaggle community competition "DL Sprint"\footnote{https://www.kaggle.com/competitions/dlsprint}, under the name "YellowKing". Among 59 participating teams, our model was the best performing one on a hidden dataset, achieving a Levenshtein Distance of 6.234\footnote{https://www.kaggle.com/competitions/dlsprint/discussion/349991}, which was 1.1049 units lower than other competing submissions. Overall, this training pipeline generalized well on completely unseen data compared to other models. Our submission also achieved the highest combined score of 92.59 out of 100, taking the public dataset, hidden dataset and methodology into account. 

\subsection{Base Model}
Most of the contenders chose \textit{arijitx/wav2vec2-xls-r-300m-bengali} as the model to start training. That model was already fine-tuned for 180,000 steps, on the OPENSLR-SLR53-Bengali dataset\cite{kjartansson-etal-sltu2018}. The rationale for this choice was to utilize the already learned weights from that model, to incorporate transfer learning. This choice allowed such pipelines to quickly achieve a low Levenshtein Distance and fit the train-validation set better than our pipeline\footnote{https://www.kaggle.com/competitions/dlsprint/leaderboard}.

While Bengali Common Voice Speech Dataset and OpenSLR-Bengali are datasets of the same language, the Bengali Common Voice Speech Dataset has higher median voice segments per second, higher dynamic range, smaller pauses between voice segments\cite{cvbn}. The Bengali Common Voice Speech Dataset is crowd sourced and therefore recorded in a more uncontrolled environment. In addition, OpenSLR-Bengali was recorded by a total of 505 speakers, whereas Bengali Common Voice Speech Dataset has about 20,000 speakers, having accents originating from different parts of Bangladesh and India. This allows models trained on that dataset to perform better on unseen practical data with a varying range of speaking patterns. 

We chose to start training \textit{facebook/wav2vec2-large-xlsr-53}, which was not fine tuned for Bengali speech transcription. We trained it solely on the Bengali Common Voice Speech Dataset, which allowed it to be exposed to data from a more uncontrolled environment and varying accents from the beginning. Therefore, our model slowly improved it's Levenshtein Distance over time, as opposed to the rapid learning rate observed on the contender pipelines, but performed better on unseen practical data. 

\subsection{Filtering Data}
All of the contenders trained their model on almost the entirety of the \textit{Bengali Common Voice Speech Dataset}\cite{cvbn} on about 200,000 samples. The natural reasoning for this is, increasing the number of samples allows the model to be trained on more rich corpus of words and sentences. 

The dataset was divided on the basis on manual verification, using "up vote" and "down vote" metrics. Which allowed us to filter the verified data from the unverified ones. We chose to train on a small subset of the data, including on the samples that had $upvotes > downvotes$. Upon random sampling of the data excluded from our training set we observed that their flaws were - stuttering, restarting sentence upon mistake, mispronunciation. Therefore we chose to train on a cleaner dataset, consisting of about 37,500 training samples, and 7,747 validation samples.

\subsection{Training Sequence}
Most of the contenders trained their models on the bulk 200,000 samples. This data includes the errors mentioned earlier. Therefore their models learned to capture the ambiguous pronunciations, sentence structures and unnatural pauses. 

Whereas we chose to train our model on two phases. First phase included the majority of our training set consisting of 37,500 samples. Then later in the training process, we combined the training and validation set to have 45,000 samples. This allowed for gradual exposure of new vocabulary to the model.

\section{Discussion}

\subsection{Training Loss and WER Correlation}
While training it was observed there that the training loss did not correspond strongly to the Word Error Rate. Since WER is the final evaluation metric, epochs in the later stages of training sometimes resulted in slight but temporary increases in training loss that may persist for multiple epochs before lowering once again. Unlike other neural network models, the risk of over-fitting appears to be low in the Transformer model.

\subsection{Effect of Exposure to New Data}
The model was also able to adapt to exposure to new data well and it did not result in any sudden spike in evaluation metrics. On the second dataset, we observed that even with training loss and WER plateauing, further training resulted in a slightly improved Levenshtein Distance. This is likely due to the exposure to new vocabulary which ultimately improves performance on previously unseen data.

\subsection{About the facebook/wav2vec2-large-xlsr-53 pretrained model}

The wav2vec2-large-xlsr-53 model which we used for further fine-tuning was pretrained on about 56 thousand hours of multilingual speech from the MLS, CommonVoice and BABEL datasets\footnote{https://github.com/facebookresearch/fairseq/tree/main/examples/wav2vec}. However the majority (50.7 thousand hours) were on the MLS dataset \cite{Pratap2020MLSAL} which comprises of Dutch, English, French, German, Italian, Polish, Portuguese and Spanish; i.e. Modern European Languages.
\linebreak
For this reason, effect of pretraining on cross-lingual speech appears somewhat muted on non-European languages. The xlsr model showed significant phonetic token error rate (PTER) improvement for Georgian \cite{xu2021simple}, a language similar to the languages it was pretrained on, but marginal improvement on Bengali and even deterioration on Vietnamese compared to the previous state-of-the-art \cite{gao2021zero} on low resource ASR.

\begin{table}[h!]
\centering
\caption{PTER on Babel Languages 6 Dataset}
    \begin{tabular}{|c|c|c|}
    \hline
       Language  & Gao et. al \cite{gao2021zero} &   wav2vec2-large-xlsr-53 \cite{xu2021simple} \\ 
       \hline
       Georgian  & 38.6 & 27.6\\
       Bengali & 38.2 & 36.1\\
       Vietnamese & 32.0 & 40.7\\
       \hline
    \end{tabular}

    \label{table:3}
\end{table}

We believe that a model pretrained on languages similar to Bengali would offer similar improvements for Bengali ASR.

\subsection{Effectiveness of Transformer Model}
It is our belief that we didn't push the limits of the wav2vec2 Transformer model since in the ideal case (pretraining, fine-tuning and inference on English), it achieved a best case WER of 0.018 \cite{yi2020applying} without using a language model. With more labelled data and better compute, it is highly likely a better result on Bengali speech recognition can be achieved using the transformer architecture.

\section{Conclusion and Future Work}

In this work, we present an effective scheme of fine-tuning a pretrained wav2vec2.0 model for speech transcription of a low resource language. We applied the scheme on a subset of 45 thousand audio samples from the Bengali Common Voice Dataset to transcribe Bengali audio, achieving a final WER of 0.2524. This demonstrates the viability of the wav2vec2 model for Bengali Speech Recognition. Furthermore, this result was achieved using only 17.84\% of the Bengali Common Voices Dataset making it very likely that even better results can be achieved if the entire dataset is utilized. Furthermore, the model was pretrained on mostly modern European languages, with very little exposure to Bengali or languages similar to it. We anticipate that pretraining the model on languages similar to the target language will yield better results as well. We will train networks using our scheme on future iterations of the dataset to create robust models.

\bibliography{refs}
\vspace{12pt}

\end{document}